\documentstyle[sprocl]{article}

\def\NCA{\em Nuovo Cimento}
\def\NPB{{\em Nucl. Phys.} B}

\def\PLB{{\em Phys. Lett.} B}
\def\PR{{\em Phys. Rev.}}

\def\PRD{{\em Phys. Rev.} D}

\def\PRL{\em Phys. Rev. Lett.}

\newcommand{\be}{\begin{equation}}
\newcommand{\ee}{\end{equation}}
\newcommand{\bea}{\begin{eqnarray}}
\newcommand{\eea}{\end{eqnarray}}

\newcommand{\nonu}{\nonumber\\}

\begin{document}

\title{ INSTANTON INDUCED PHENOMENA IN THE EFFECTIVE STANDARD MODEL}

\author{VINCENZO BRANCHINA}
\address{Laboratory of Theoretical Physics, Louis Pasteur University\\
3 rue de l'Universit\'e 67087 Strasbourg, Cedex, France\\
branchina@sbgvax.in2p3.fr}

\author{JANOS POLONYI}

\address{Laboratory of Theoretical Physics, Louis Pasteur University\\
3 rue de l'Universit\'e 67087 Strasbourg, Cedex, France\\
polonyi@fresnel.u-strasbg.fr}
\address{Department of Atomic Physics, L. E\"otv\"os University\\
Puskin u. 5-7 1088 Budapest, Hungary}
\maketitle\abstracts{
It is shown that in the Effective Standard Model new phenomena can arise due 
to the presence of {\it small} instanton configurations. The chief result is 
that under certain conditions new  {\ hidden coupling constants} could exist 
in the model. In the Electroweak sector that might result in the possibility
of observing $B+L$ violating processes due to a highly non perturbative
contribution to the {\it holy grail function}. The same phenomenon might 
occur in the $QCD$ sector of the theory and could be observed in the 
DIS experiments at HERA.}

\section{Introduction}

According to the Principle of Renormalizability all the interaction terms
whose coupling constant has negative energy dimension  
should be discarded from our fundamental Lagrangians because they give rise 
to non renormalizable theories\footnote{Non renormalizability
generally means that 
it is not possible in the framework of perturbation theory to absorb all
the infinities appearing in the computation of the physical quantities 
simply adjusting a finite number of bare parameters. Recently this
definition of renormalizability has been enlarged to include also 
theories with non renormalizable terms\cite{wein}. 
What is important for the purpose of our contribution is 
that even under this larger definition of renormalizability the analysis 
is still performed within perturbation theory.}.

A better insight came after the work of Wilson who has pioneered  
a real change in our way of conceiving renormalization.
It is no more simply regarded as a welcome technical device to 
get rid of the unwanted divergencies in our physical quantities 
but rather as the necessary manifestation of the difference in the 
description of Physical phenomena at different scales. The energy cutoff
energy also acquires a physical meaning, it represents the 
maximal scale at which the theory itself makes sense. 

In this novel perspective the non renormalizable operators acquire a different
ontological status. Consider a Langrangian containing such
operators. It may be proven  that {\it at every order in perturbation theory }
the contributions to any physical quantity coming from those terms are 
always suppressed by negative powers of the cutoff. As far as the energy scale
at which the process under consideration occurs is very far from the cutoff
scale, these contributions are practically negligible. The technical notion of 
non renormalizable operator is replaced by the more physical one of
irrelevant operator. The renormalizable operators similarly are called
relevant operators \footnote{To be precise the renormalizable operators are 
the relevant and the marginal ones.}.

The alert reader has certainly remarked that we have explicitly stressed that
the classification of operators mentioned above has been done within the 
framework of perturbation theory. 
We mention now two situations where such a classification might suffer from 
serious changes.

i) In order to study the running of the coupling constants of our model we 
generally first identify a fixed point and then linearize the evolution 
equations. Consequently this classification is
valid until we reach the edge of the linearization region. Nothing can be said
about the role of the irrelevant operators outside of this region. 
In this volume an explicit example is given of such a violation of universality
behaviour together with the possible physical consequences of such a 
phenomenon\cite{jean}.

ii) Another possible source of violation of the universality behaviour is 
related to the presence of non trivial saddle points. 
In the path integral language the perturbative analysis mentioned above  
corresponds
to a semiclassical approximation in the presence of trivial saddle points.
If the action under investigation possess non trivial saddle points, due to 
this very non perturbative effect the above classification might fail. 
The purpose of this contribution is to give an explicit example of such a
failure together with the possible
interesting physical consequences associated with it.

\noindent
In Sec. 2 we review and partially extend the results already 
presented\cite{nouniv} on the analysis of an effective $SU(2)$ model.
In Sec. 3 we discuss the instanton induced effects in the light of our
results.

\section{ SU(2) in the presence of irrelevant operators}

We believe that the strongest argument which suggests that these kind of
models really deserve our attention comes once again from the Wilsonian
approach to field theory. 
Any model but the Theory of Everything is the low energy approximation of 
some more fundamental one. In the process of eliminating higher energy 
degrees of freedom, irrelevant terms appear in our effective Lagrangians.

Let's consider an SU(2) Lagrangian containing all sort of irrelevant terms
as for instance higher power or higher derivative terms
\be
L_{eff}=-{1\over4g^2}G^a_{\mu\nu}\biggl( 1 +
\sum_n {c_{2n}\over\Lambda^{2n}}(D^2)^n +
\sum_m {d_{2m}\over\Lambda^{2m}}(G^b_{\alpha\beta}G^{b\alpha\beta})^m
+ \cdots \biggr)G^{a\mu\nu}.
\label{high}
\ee
$\Lambda$ is the cutoff of the theory and $D$ is the covariant derivative.

\noindent
Due to the topology of the configuration space the path integral defining the
theory has saddle points which become the self dual instantons for 
$c_n = d_m=\cdots = 0$. 
We now introduce the collective symbol ~{\bf c}~ to indicate the whole set of 
irrelevant coupling constants and denote such a distorted instanton, 
~~ $A_\mu^{({\bf c},\rho)}(x).$~~
$\rho$ is the scale parameter of the instanton configuration, 
which is introduced in a somehow arbitrary 
fashion by requiring that the distorted instanton be a self dual instanton with
size $\rho$ for ~{\bf c}=0. 

\noindent
The tree level action for this configuration is of the 
form 

\be
S_{\rm inst}(\rho)=
{8\pi^2\over g^2}
\biggl(1- f({\bf c}, \rho\Lambda) \biggr)
\label{confa}
\ee
where ${8\pi^2\over g^2}$ is the scale independent usual instanton action.
The dimensionless function ~$f$~ is a complicated function of
the whole set of irrelevant coupling constants 
~~{\bf c}~ as well as of the dimensionless product $\rho\Lambda$.
Depending on the actual value of the coupling constants ~~{\bf c}~~
the function $S_{\rm inst}(\rho)$ as a function of $\rho$
may or may not have a minimum 
for a value $\bar\rho$ of $\rho$ which, for dimensional reasons is 
$\bar\rho\sim {1\over\Lambda}$.
When there is no minimum no surprises occur in the model.
Let's then analyze the case when such a minimum exists. In particular the
value of ~~$S_{\rm inst}(\bar\rho)$~~ at the minimum can be positive or 
negative. When ~~$S(\bar\rho)$~~ is negative new interesting phenomena 
occur\cite{jochen}\cite{herve}
Here we limit our analysis to the case ~~$S_{\rm inst}(\bar\rho) > 0$.~~

\noindent
Let's call ~~$h$~~ the value of ~~$f$~~ at the minimum. We have
\be
S_{\rm inst}(\bar\rho)=
{8\pi^2\over g^2}
\biggl(1- h\biggr)
\label{minn}
\ee
Computing now the contribution to the partition function coming from the 
instanton sector, $Z_1$,~ normalized as usual with the contribution from the
zero winding number sector, $Z_0$,~ we get\cite{nouniv} 
\be
\biggl({Z_1\over Z_0}\biggr)=C{V\over g^8}\int_0^{1\over\mu}
{d\rho\over\rho^5}e^{-{S_0\over g^2}\bigl(1- f({\bf c},g, \rho\Lambda)
\bigr)}{\cal D}(\rho\Lambda)
\label{ratio}
\ee
where 
${\cal D}$ stands for the contribution of the fluctuation determinant,
$V$ is the quantization volume and $C$ is a numerical constant whose value is 
of no importance for us.
The infrared catastrophe is ignored by the introduction of the infrared
cut-off, $\mu$~ which may be regarded as the $\Lambda$  
parameter of the theory (say $\Lambda_{QCD}$ in $QCD$).

\noindent
The contribution of instantons as the function of the size parameter
has a well pronounced peak at $\rho\approx\bar\rho$. Another important region
is in the infrared where the loop corrections increase.
In order to separate the contribution of the large, i.e. cut-off
independent instantons from that of the stable saddle point in the
vicinity of the cut-off we split the scale integration into two parts
by the help of a scale parameter $m$, $\Lambda_{QCD}<<m<<\Lambda$,
\be
\int_0^{1\over\mu}d\rho\cdots=\int_0^{1\over m}d\rho\cdots+
\int_{1\over m}^{1\over\mu}d\rho\cdots.
\label{intspl}
\ee
The first and the second integral will be referred as the contribution 
of the mini and the large instantons and denoted as
$(Z_1/Z_0)_L$ and $(Z_1/Z_0)_M$ respectively. 
The final result is 
\bea
\biggl({Z_1\over Z_0}\biggr)_L&=&
C{V\over g^8}\int_{1\over m}^{1\over\mu}{d\rho\over\rho^5}e^{-{8\pi^2\over g^2}
\bigl(1- f \bigr)}
{\cal D}(\rho\Lambda)
\sim{3\over10}C{V\Lambda^4\over g^8}
e^{-{8\pi^2\over g^2}}\biggr({\Lambda\over\mu}\biggl)^{10\over3}\nonu
\biggl({Z_1\over Z_0}\biggr)_M&=&C{V\over g^8}\int_0^{1\over m}
{d\rho\over\rho^5}e^{-{8\pi^2\over g^2}
\bigl(1- f \bigr)}
{\cal D}(\rho\Lambda)
\sim C{V\Lambda^4\over g^8}e^{-{8\pi^2\over g^2}(1 - h)}
{\cal D}(\bar\rho\Lambda)
\label{lm}
\eea

We now compare the contributions coming from these two different regions
by computing the ratio
\be
R = {\bigl({Z_1/Z_0}\bigr)_{L}\over\bigl({Z_1/Z_0}\bigr)_{M}}
\sim\biggl({\Lambda\over\mu}\biggr)^{{2\over3}(5 - 11h)}
\label{rat}
\ee
The result here is that for ~~$h > {5\over11}$~~
the {\it mini-instantons} dominate the path integral, which is otherwise 
dominated by the large instantons.

What are the consequences of such a result?

\section { Coupling constants flow, a new relevant parameter}

Suppose that we are in the case when the large instantons dominate the path
integral (i.e. $h < {5\over 11}$). 
In order to extract the flow of the coupling constant ~$g$~ 
we demand as usual\cite{thooft}
the cutoff independence of ~$\bigl(Z_1/Z_0\bigr)_L$~, 
i.e. we require that
~~$ \Lambda{d\over d\Lambda}\bigl({Z_1/ Z_0}\bigr)_L=0$.
We immediately obtain
\be
\Lambda{d\over d\Lambda}\biggl(\Lambda^{22\over3}
e^{-{8\pi^2\over g^2}}\biggr)=0 
~~~~\to~~~~\beta(g)=-{11\over24\pi^2}g^3
\label{betal}
\ee
i.e. we get the same result of the perturbative analysis in the zero winding 
number sector. This is the well known t'Hooft result. It is by 
no means surprising that we get the same result as in the perturbative 
constant background, since the ultraviolet structure is the same
on the flat or the large instanton background. By the way
the large instantons are the only ones considered in the standard
analysis\cite{thooft}. 

If we now apply the same strategy to extract the beta function for ~$g$~ 
in the case when the mini-instantons dominate the path integral, i.e. we 
require that 
~~$ \Lambda{d\over d\Lambda}\bigl({Z_1/ Z_0}\bigr)_M=0$, we find
\be
\Lambda{d\over d\Lambda}\biggl(\Lambda^4
e^{-{8\pi^2\over g^2(1 - h)}}\biggr)=0
~~~~\to~~~~ \beta(g)=-{1\over4\pi^2(1 - h)}g^3
\label{betam}
\ee
This is a very surprising and disturbing result. It seems we have run into
a paradox. In fact in a theory as $SU(2)$ we should distinguish between two
different classes of observables. Those with a topological origin,
as the topological susceptivity, 
the mass of the $\eta^{'}$, certain massless Green functions,
and observables with non topological origin, as for instance the
Wilson loop or the cross section for an ordinary process.
The renormalization of those observables insensitive to the 
topology, whatever is the value of ~$h$,~ requires the
beta function for ~$g$~ dictated by the perturbative analysis. 
On the other hand our result shows that the observables with topological origin 
are sensitive to the actual value of ~$h$.~For ~$h > 5/11$~ it seems 
that their renormalization requires a different
flow for ~$g$,~ given by eq. (\ref{betam}). Is there any 
possibility to reconcile these results?

The very origin of this apparent contradiction lies in our conservative 
attitude to require a fine tuning only for the parameter ~$g$~ which we 
consider to be the only relevant parameter of the theory. For the parameter 
~$h$,~ which is the result of an intricate combination of all the irrelevant 
parameters of our effective theory, we do not require any fine tuning, 
according with the statement that it is irrelevant. 
We shouldn't forget however that its appearance
is related to the presence of non trivial saddle points in 
our model. As we have already mentioned in the Introduction 
there is no guarantee that the perturbative classification of the
operators should still be valid in this case. 
Actually we may easily convince ourselves that this is indeed not the case.
Due to the presence of such irrelevant operators, with dimensionful coupling 
constants containing negative cutoff powers, the theory requires a sort of
renormalization at the tree level, a phenomenon which cannot be 
accounted for by the power counting theorem,  which is based on the 
analysis of the loop corrections only.

The resolution of this apparent paradox is very simple. In eq.
(\ref{betam}) we have actually another parameter,~$h$,~ which can be fine
tuned.  
We can now keep the ratio ~$\bigr(Z_1/Z_0\bigl)_M$ finite and maintain the 
usual flow for ~$g$,~ fine tuning appropriately the coupling constant ~$h$.~ 
In other words, the two conditions
\be
\Lambda{d\over d\Lambda}
\biggl(\Lambda^4 e^{-{8\pi^2\over g^2(1 - h)}}\biggr) = 0
~~~~~~~~{\rm and} ~~~~~~~~
{8\pi^2\over g^2}={22\over3}{\rm ln}{\Lambda\over\Lambda_0},
\label{condi}
\ee
where $\Lambda_0$ is the $\Lambda$- parameter of the theory,
give
\be
h(\Lambda)= {5\over11} + {\kappa\over{\rm ln}(\Lambda/\Lambda_0)}.
\label{new}
\ee
$h$~ is actually a {\it relevant parameter} of the theory.
This is the very crucial and new result of our analysis.
The theory posses two different phases. In the mini instantons dominated phase
there is a new {\it hidden} relevant coupling constant and its flow is given by
eq. (\ref{new}). Note that the actual value of this new coupling constant
at a given energy scale is a new external input which can be determined 
only via experiments, precisely the same way as the electron
mass and charge in QED are determined by the experiments and are the external 
input of the theory. 

We may now rewrite the expression for $\bigr(Z_1/Z_0\bigl)_M$ by the help of 
a physical finite scale $\mu$ in the following manner
\be
\Lambda^4 e^{-{8\pi^2\over g^2(\Lambda)}\bigl(1 - h(\Lambda)\bigr)} = 
\mu^4     e^{-{8\pi^2\over g^2(\mu    )}\bigl(1 - h(\mu    )\bigr)} =
\mu^4     e^{-{8\pi^2\over g^2(\mu    )}\bigl({6\over11} - 
{\kappa\over{\rm ln}(\mu/\Lambda_0)}\bigr)}
\label{fine}
\ee
This last expression will be helpful to understand 
the physical implications of our result.

\section{Instanton induced effects }

Since the pioneering work of t'Hooft\cite{thooft}, it is well known that 
~$B+L$~ is not strictly conserved in the Standard Model.
Certain massless Green functions which vanish in perturbation
theory, get a non perturbative contribution from the
instanton sector.
This effect is due to the anomalously non conserved current\cite{jack},
but the small 
instanton suppression factor ~~$exp(-{4\pi\over\alpha_w})$~~
makes it unobservable. 
At the beginning of the ninethies there was a certain 
excitement due to the discovery\cite{ring} of an enhancement factor for 
the $B+L$ violating cross sections.
It was found that for those $B+L$ violating processes 
with high multiplicity of ~~$W, Z$ , and/or ~$H$~~ particles in the final state,
the cross section get enhanced w.r.t. the t'Hooft suppression factor, 
growing with the center of mass energy, $s$, as
\be
\sigma\sim {\rm exp}\Biggl(-{4\pi\over\alpha_w}
\Bigl(1-{\rm const}\times
\bigl({s\over M_{sph}^2}\bigr)^{2\over3}\Bigr)\Biggr).
\label{sect}
\ee
$M_s$ is the sphaleron\cite{manton} mass.
More detailed studies have subsequently shown that the approximations 
involved in 
those original computations loose their validity at energies far below the 
scale $M_s$ 
where the whole argument of the exponential, the so called 
{\it holy-grail function}, would become small.
The final agreement was that at best the suppression exponent 
could reach a value of one half the original t'Hooft factor and then
the effect would be still far beyond any possibility of observation. 

Our result gives a novel mechanism
which allows the {\it holy-grail function} to receive an O(1) contribution
( see the last expression of eq.(\ref{fine}))
with an opposite sign with respect to the t'Hooft suppression factor,
independently of any enhancement due to the high multiplicity of the
final state. It opens again the door for the possible observation of $B+L$
violating processes within the framework of the Standard Model at 
accessible energies. 

There is also another interesting possible application of the results
presented above. During the last years, even though the scientific community 
has somehow given up the hope to observe instanton induced $B+L$ violating
processes in the Standard Model, there has been an intense theoretical
activity\cite{{ringy},{baty}}
concerning the possible impact of instanton induced processes in
QCD, particularly in DIS processes. The basic mechanism is actually the same,
the only difference being that it has to be applied to $SU(3)$ rather then
to $SU(2)$. This field is actually under active investigation and a dedicated 
working group\cite{ringy} exists in connection with the HERA experiments.
The hope is to finally observe the
impact of the presence of instantons in the cross sections for these processes.
These experiments may turn out to be very important from the point of view
of our result. In addition to the usual well studied contribution to the 
{\it holy grail function} our result suggests that
an additional contribution might come from the presence of an 
{\ it hidden} coupling constant in $QCD$. 

\noindent
The comparison of the experimental cross sections with the theoretical 
predictions might eventually allow to measure the {\it new scale} introduced
in the theory through the presence of the constant $\kappa$ in eq.(\ref{new}).


\begin{thebibliography}{100}
\bibitem{wein} S. Weinberg {\PRD {\bf 56}} (1997) 2303;
J. Gomis and S. Weinberg {\NPB {\bf 469}} (1996) 473.
\bibitem{jean} J. Alexandre, V. Branchina, J. Polonyi, 
{\it Renormalization group with condensate}, these Proceedings.
\bibitem{nouniv} V. Branchina, J. Polonyi, {\it Mini-Instantons in 
$SU(2)$ Gauge theory}, hep-th/9606160.
\bibitem{jochen} J. Fingberg and J. Polonyi, {\it
Anti-Ferromagnetic Condensate in Yang-Mills Theory},
hep-lat/9602003, to be published in \NPB.
\bibitem{herve} V. Branchina, H. Mohrbach, J. Polonyi, 
{\em The Antiferromagnetic $\phi^4$ model, I., II.}, 
hep-th/9712110,9712111, submitted to {\em Nucl. Phys.}
\bibitem{thooft} G. t'Hooft, {\PRL {\bf 37}} (1976) 8;
G. t'Hooft, {\PRD {\bf 14}} (1976) 3432.
\bibitem{jack} S. Adler, {\PR {\bf 177}} (1969) 2426;
J. Bell and R.Jackiw, {\NCA ~{\bf 60A}} (1969) 47. 
\bibitem{ring} A. Ringwald, {\NPB {\bf 330}},1 (1990), 1;
O. Espinosa, {\NPB {\bf 343}}, 310 (1990).
\bibitem{manton} F. Klinkhamer, N. Manton, {\PRD {\bf 30}}(1984) 2212.
\bibitem{ringy} 
A. Ringwald and F. Schrempp, {\it Instanton Phenomenology at
HERA}, hep-ph/9706399;
A. Ringwald and F. Schrempp, {\it Instanton-Induced Processes in
Deep-Inelastic Scattering}, hep-ph/9610213.
\bibitem{baty} I. Balitsky and V. Braun, {\PLB {\bf314}} (1993) 237.
\end{thebibliography}
\end{document}